\documentclass{pazh_engl}

\usepackage{epsf}
\usepackage{flushrt}
\usepackage{graphicx}

\usepackage{pstricks}
\begin{document}

\sloppypar

\title{\bf IGR J16318-4848: an X-ray source in a dense envelope?}

\author{\copyright 2003 M. Revnivtsev\inst{1,2},
S. Sazonov\inst{1,2}, M. Gilfanov\inst{1,2}, R. Sunyaev\inst{1,2}}

\institute{Max-Planck Institute for Astrophysics, Garching, Germany
\and
Space Research Institute, Russain Academy of Sciences, Moscow, Russia}

\authorrunning{Revnivtsev et al.}
\titlerunning{}
\date{}
\abstract{
The hard X-ray source IGR J16318-4848 was recently
discovered by the INTEGRAL observatory (Courvoisier et al.) and
subsequently uncovered in archival data of ASCA observations in 1994
(Murakami et al.). We present results of a detailed analysis of the
ASCA data. The spectrum of the source in the 0.5--10 keV band is
extraordinarily hard and is virtually unobservable below 4 keV because
of strong photoabsorption ($n_H L>4\cdot 10^{23}$ cm$^{-2}$). The
4--10 keV emission is dominated by a $K_\alpha$ line of neutral or weakly ionized iron 
with an equivalent width of $\sim 2.5$ keV. There is also an
indication for a second line at $\sim$ 7 keV. Our analysis of archival
observations of the IGR J16318-4848 infrared counterpart, discovered
by Foschini et al., shows that the point source is detected at
different wavelengths in the 1--15 $\mu$m range. The available
data suggest that IGR J16318-4848 is an X-ray binary system
enshrouded by a dense envelope. It is possible that the source is a
wind-fed high-mass X-ray binary similar to GX 301-2. We argue that IGR
J16318-4848 might be the first representative of a previously unknown
population of highly absorbed galactic X-ray sources, which remained
undetected with X-ray missions before INTEGRAL.
}
\titlerunning{IGR J16318-4848: an X-ray source in a dense envelope?}
\maketitle

\section*{Introduction}

The launch of the powerful hard X-ray/gamma-ray observatory INTEGRAL 
(http://astro.estec.esa.nl/SA-general/Projects/Integral/integral.html) 
will possibly lead to the discovery of a new population of sources
highly absorbed below $\sim 10$ keV, which could have been missed by
previous X-ray missions. 

It has remained a puzzle for many years why there
are no X-ray sources in rich molecular clouds, where intense star
formation is known to be taking place and where massive binaries
should be present in principle. Another reason for us to expect
the existence of X-ray sources in molecular clouds is connected with
the possibility of a passage through them of numerous single and binary
neutron stars ($\sim 10^{9}$ per galaxy) and black holes ($\sim
10^{7}$--$10^{8}$ per galaxy). Simple estimates using the Bondi
formula indicate that a significant brightening of such an
object could result from accretion of the surrounding molecular
cloud. However, the large optical depth to absorption of massive
molecular clouds can considerably diminish the outgoing X-ray
radiation and make such sources practically unobservable for X-ray
telescopes operating in the soft or standard X-ray bands as well as
for all-sky monitors, most of which also work in these bands. 

From another point of view, super-Eddington accretion, widely discussed
by theorists, could also lead to almost total obscuration of X-ray
sources below $\sim 10$ keV. Two basic scenarios
of supercritical accretion are being considered: the formation of
a geometrically and optically thick accretion disk, which carries
trapped radiation into the black hole (Abramowicz et al. 1988), or 
the appearance of a strong outflow of matter from an accreting black
hole or a neutron star (Shakura \& Sunyaev 1973). It is possible that
at a certain stage, binaries like SS 433 behave in this way and we
have observed the same regime during an outburst of V4641
Sgr, where a dense, absorbing atmosphere was formed around the
accreting object (Revnivtsev et al. 2002). 

Yet another possible class of objects that could have remained
undetected with X-ray observatories before INTEGRAL is binary X-ray
sources, including X-ray pulsars, embedded in dense wind material from
a donor star. Such an envelope could render the X-ray source
almost invisible at energies below $\sim 5$ keV due to
photoabsorption. Moreover, the envelope could become Compton thick in
extreme cases, leading to a downscattering of 
hard X-ray photons to lower energies and the formation of an emergent
spectrum with a peak at 10--20 keV.

The discovery (Courvoisier et al. 2003) of the strongly absorbed
source IGR J16318-4848 by the INTEGRAL observatory is
significant. Thanks to its high sensitivity at energies above 15 keV,
INTEGRAL is capable of discovering relatively weak, highly absorbed
X-ray sources of the types described above.

The INTEGRAL observatory spends a considerable amount of time scanning
the Galactic plane. During one of such scanning observations, on
Jan. 29, 2003 a previously uknown source was discovered. It was 
named IGR J16318-4848. According to the data of the IBIS telescope 
(operating in the energy band 15 keV-10 MeV) the position of the
source was determined to be: R.A. $=16^{\rm h}31^{\rm m}52^{\rm s}$, 
Dec $=-48^{\circ}48'.5$ (equinox 2000, position uncertainty 2', Courvoisier 
et al. 2003). The measured flux from the source in the 15--40 energy
band was at a level of 50--100 mCrab and varied on a time scale of hours.

An analysis of archival data of observations of this field with the
ASCA observatory (on Sept. 3, 1994) revealed the presence of a weak
source at the position coincident with the INTEGRAL localization of
IGR J16318-4848 (Murakami et al. 2003). The authors pointed out that the
spectrum of the source was strongly absorbed and also mentioned the possible
presence of a 6.4 keV iron line. Subsequent observations of the source
by the XMM-Newton observatory showed the presence of strong iron lines
at $\sim$6.4 keV and $\sim$7.0 keV (Schartel et al. 2003, de Plaa et
al. 2003). An analysis of archival observations of the
field of the source showed the presence of an infrared counterpart at R.A. =
16$^h$31$^h$48$^s$.3, Dec. = -48$^{\circ}$49'01" (equinox 2000,
position uncertainty 0.2'', Foschini et al. 2003)

In this paper we present the results of a detailed analysis of
the ASCA observations of IGR J16318-4848. We discuss the possible nature of the
source, considering all available data on the source in different energy 
bands, part of which is published here for the first time.

\section*{Analysis of the ASCA observation}

\begin{figure}
\includegraphics[width=\columnwidth]{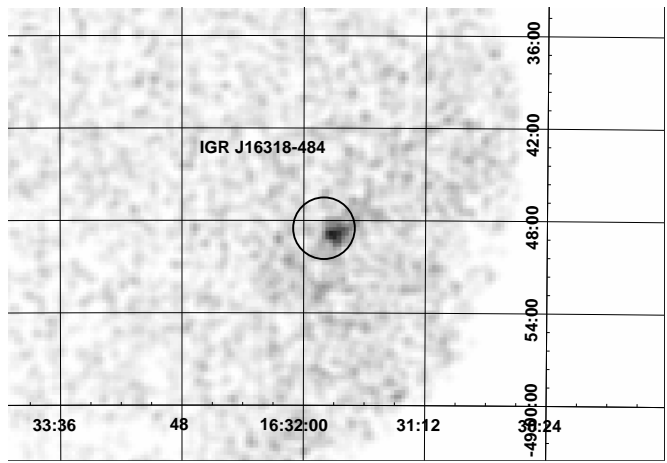}
\caption{Image of the sky around IGR J16318-4848 obtained by the ASCA
observatory.
\label{map}}
\end{figure}

\begin{figure}
\includegraphics[width=\columnwidth]{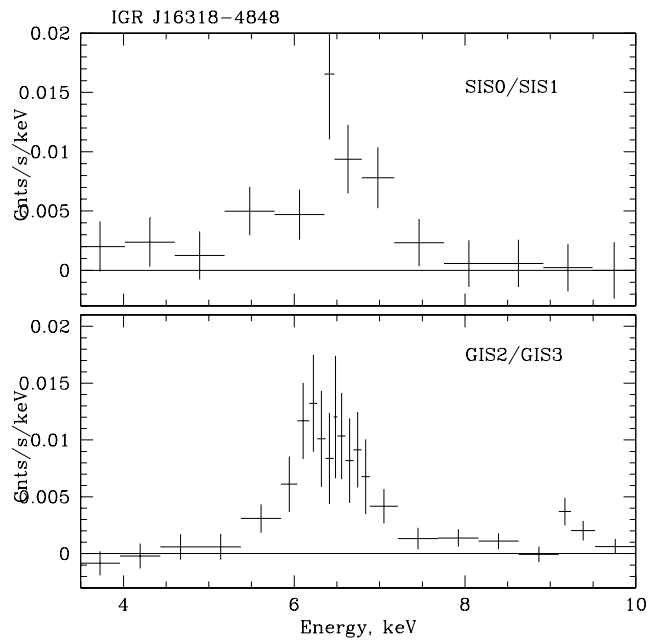}
\caption{Spectrum of IGR J16318-4848 obtained with the GIS and SIS 
spectrometers at 3.5--10 keV. Note that to derive these spectra 
we used background \#2, which has lowered the statistical quality of
the spectra (cp. Fig. 3) \label{spectrum_cnts}}
\end{figure}

A region of the sky including the position of IGR J16318-4848 was
observed by ASCA (Tanaka et al. 1994) on Sept. 3, 1994
4:11UT--7:11UT. The total exposure time was $\sim$4 ksec. In our
analysis, we used the data of all instruments of the observatory -- 
the gas spectrometers GIS and solid state spectrometers SIS. For the
data reduction we used the standard software package LHEASOFT 5.2. For
the spectral spproximation the XSPEC package was used. In order to
avoid biased best fit parameters in modeling the source spectrum at
a low source count rate, we applied the Churazov et al. (1996) method of
weighting the spectral points. 

A clear sign of stray light from the near and relatively bright source 
4U1624-49 is seen on ASCA detectors (the flux of 4U1624-49 is
approximately 50 mCrab and its distance from the center of the ASCA
field of view is approximately 50'). Because of this, in order to
avoid a 4U1624-49 contribution to the spectrum measured from IGR
J16318-4848, we used two types of background. Background \#1 was taken
from standard empty field observations and background \#2 from an 
annulus around the source with an inner radius of 4' and an outer
radius of 8'. Note that in the latter case the statistical quality of 
the IGR J16318-4848 spectrum is worse because the exposure time of the
background spectrum \#2 is much lower than that for background \#1.

A GIS image of the sky around IGR J16318-4848 smoothed with a 0.5' gaussian
is presented in Fig. 1. The circle denotes the uncertainty of the
source localization by INTEGRAL/IBIS. A single source inside this circle 
is clearly seen. Its coordinates are ${\rm R.A.} = 16^{h}31^{m}49^{s}, 
{\rm Dec.} = -48^{\circ}49'.2$ (equinox 2000, position uncertainty 0'.8; 
Murakami et al. 2003).

For the study of a broadband (0.5--10 keV) spectrum of IGR J16318-4848, we used
background \#2, because at low energies ($<$3--4 keV) the stray light
contribution becomes very important. In Fig. 2 we present spectra of 
IGR J16318-4848 obtained by GIS and SIS. It can be seen that a strong 6.4 keV 
line is present in the spectrum, and that the source is barely
detectable below 4 keV.

Let us first consider the region of the Fe line (5--9 keV). For this
purpose it is better to use SIS data, because the SIS detectors
have a much higher energy resolution that GIS. In this energy band, the
stray light contribution becomes very small and we can use background
\#1, which allows us to maximize the statistical quality of the data.

\begin{figure}
\includegraphics[width=\columnwidth]{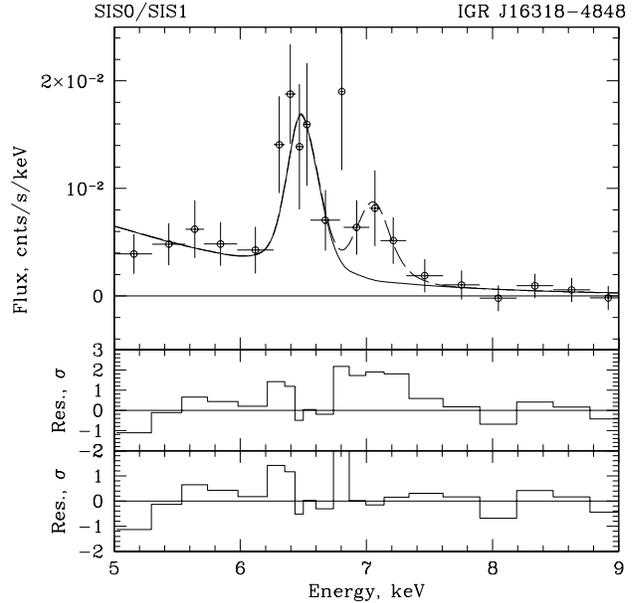}
\caption{a) Spectrum of IGR J16318-4848 according to combined data of 
SIS0 and SIS1. The solid and dashed lines represent models with one and two
emission lines. b) and c) Residuals between the observed points and the
model that includes one emission line (b) and two lines (c).
\label{twolines}} 
\end{figure}

We approximated the spectrum in the 5--9 keV energy band (Fig. \ref
{twolines}) by a power law model, $F\propto E^{-\alpha}dE$, with a
gaussian line at the energy $\sim$6.4 keV ($K_\alpha$ line of 
neutral iron). Because our spectrum has quite poor statistical quality
and we perform the spectral approximation in a very narrow energy
band, the value of the power-law index does not play an important role.
Therefore, we fixed this value at $\alpha=1.0$. The fitting 
(using the weighting method of Churazov et al. 1996)
yields a very good value of $\chi^2$=185.9/339 dof. However, there
still remain some residuals at energies around $\sim$7.0 keV, which may hint at
the presence of an additional emission line. Inclusion in the model of
a $K_\beta$ line of neutral iron with the centroid energy and width fixed
(this means an addition of only one parameter to the model) results in
an improvement of $\chi^2$ value by $\Delta\chi^2\sim$12. From the
statistical point of view, such reduction of the $\chi^2$ value means
that the inclusion of the additional parameter in the model is needed
at the false alarm probability level of $\sim 10^{-5}$. 

Therefore, there is a strong indication that the SIS
spectrum of IGR J16318-4848 contains two lines at energies
$6.46\pm0.02(\pm0.06)$keV and 7.05 keV (the value in the brackets
denotes tha systematic uncertainty of ASCA energy scale
calibration). The equivalent widths of these lines are $2.5\pm0.5$ keV 
and $2.0\pm0.6$ keV, respectively. The ratio of the fluxes of $K_\alpha$ 
and $K_\beta$ lines is $F_\alpha/F_\beta=2.4\pm0.8$ without the correction 
for photoabsorption and $F_\alpha/F_\beta=3.6\pm1.5$ taking into
account the photoabsorption ($n_HL=9.3\cdot10^{24}$ cm$^{-2}$, see 
Table 1). We note that the theoretically predicted ratio of the
fluorescent $K_\alpha$ and $K_\beta$ yields is higher, $\sim
8$. Part of the difference is probably caused by the radiative transfer in the
dense medium obscuring the X-ray source. 

Note that the presence of the second emission line at energy $\sim$7 keV
is confirmed by the results of XMM-Newton observations (Schartel et
al. 2003, de Plaa et al. 2003)

By including in the model neutral photoabsorption and fixing the 
parameters of the emission lines, we can now perform a
fit of the whole X-ray spectrum obtained using data of all ASCA
instruments. The resulting best fit values of the parameters are
presented in Table 1. 

\begin{table}
\caption{Best fit parameters of spectral approximation of IGR
J16318-4848 (ASCA data)} 
\begin{tabular}{l|c}
Parameter\\
\hline
Line energy $E_{1}$, keV&$6.46\pm0.02^a\pm0.06^b$\\
Line width $\sigma_1$, keV&$<0.2$\\
Line equivalent width $EW_1$, keV$2.5\pm0.5$\\
Line flux $F_1$, $10^{-4}$ phot/s/cm$^2$&$9.0\pm2.1$\\
Line energy $E_{2}$, keV&7.05$^c$\\
Line width $\sigma_2$, keV&0.0$^c$\\
Line equivalent width$EW_2$, keV&$2.0\pm0.6$\\
Line flux $F_2$, $10^{-4}$ phot/s/cm$^2$&$3.7\pm1.6$\\
\hline
Photon index&1.0$^?$\\
$n_H L, 10^{22}$cm$^{-2}$&$94\pm35$\\
FLux (2--10 keV)$10^{-11}$erg/s/cm$^2$&$1.2\pm0.2$\\
$\chi^2$/dof&557.9/882\\
\hline
\end{tabular}

$^a$ - statistical uncertainty

$^b$ - systematic uncertainty

$^c$ - the quantity was fixed at this value
\end{table}

\begin{figure}
\includegraphics[width=\columnwidth]{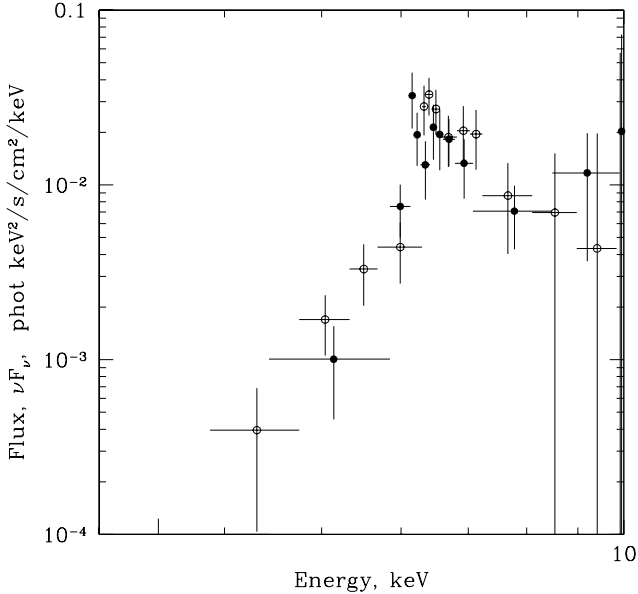}
\caption{Spectrum of IGR J16318-4848 according to the SIS 
(open circles) and GIS (filled circles) data.
\label{phot}}
\end{figure}

\section*{Archival observations of the source in different energy bands}

The X-ray coded-mask telescopes MIR/KVANT/TTM (Brinkman et al. 1985) and
GRANAT/SIGMA (Paul et al. 1991) observed the sky region around IGR J16318-4848
several times (SIGMA observations of this field are described e.g. in
Trudolyubov et al. 1998), however the source was never detected by these
instruments. Given the sensitivity of these telecopes, we can thus put
upper limits on the average X-ray flux from the source: $F_{2-30\,\,
keV} <2\cdot 10^{-10}$ erg/s/cm$^2$ (TTM) and $F_{40-100\,\,
keV}<10^{-10}$ erg/s/cm$^2$ (SIGMA). The sensitivity of these
telescopes to short (with a duration of the order of a day) flares is
worse -- the peak flux of the source during such a flare could not
have been higher than $5\cdot 10^{-10}$ and $2\cdot 10^{-10}$
ergs/s/cm$^2$, respectively. Comparison of the GRANAT/SIGMA
upper limit with the flux detected by INTEGRAL/IBIS gives an
indication that the source could be variable at least at time scale
of years. However, we should note that the energy band mentioned by 
INTEGRAL/IBIS (Courvoisier et al. 2003) does not overlap with energy 
band of GRANAT/SIGMA (40 keV - 1 MeV).

An analysis of archival observations of the field of the source made
by Foschini et al. (2003) showed the presence of an infrared
counterpart in several spectral bands (J, H, K, 8.6 $\mu$m), and also
made it possible to place an upper limit on the source brightness in
the $R$ band $m_R>21$. We reanalized the data of the 2MASS infrared survey
and obtained the following magnitudes: $m_J\sim10.2$, $m_H\sim8.5$,
$m_K\sim7.5$. Note, that the data of 2MASS survey in the sky region at
hand do not, at the present time, allow one to perform precise
photometric measurements. Therefore, the values above should be
regarded as rough estimates of the real brightness values.

Our analysis of the DSS2 survey (http://archive.eso.org/dss/dss/) and
the CAI/MAMA survey (http://dsmama.obspm.fr/) clearly showed the presence of
the counterpart in the $I$ spectral band with $m_I\sim15$ and a weak
detection in the $R$ spectral band with $m_R\sim19-20$.

Finally, our analysis of data of the MSX Galactic plane survey 
(Price et al. 2001) showed the presence of the IGR J16318-4848
counterpart at 11--13 $\mu$m ($\sim$0.54 Jy), 13--16 $\mu$m
($\sim$0.44 Jy) and a possible detection at 18--26$\mu$m ($\sim$0.4 Jy).

All the flux values of IGR J16318-4848 discussed above, obtained at
different times and in different band, are collected together in Fig.5.

\begin{figure}
\includegraphics[width=\columnwidth]{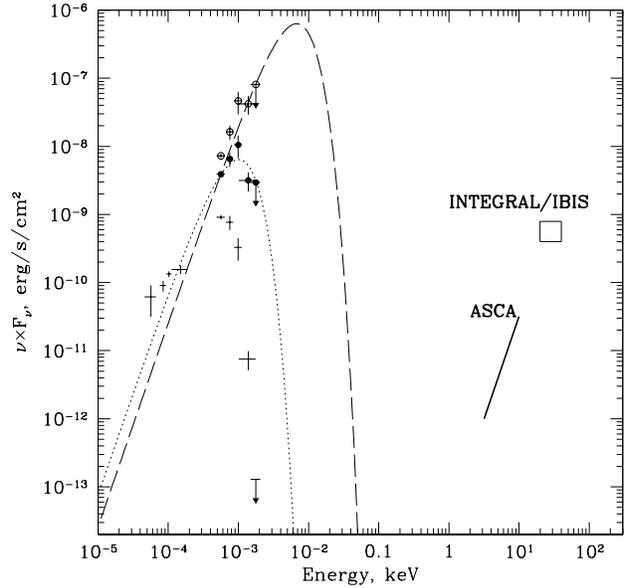}
\caption{Broadband spectrum of IGR J16318-4848. Crosses represent the 
measured fluxes of the source in infrared bands. Filled circles
represent the source flux, corrected for the galactic reddening with 
$A_v=13$, open circles with $A_v=20$. The dotted and dashed curves 
represent a rough approximation of the near infrared points by a
Planck spectrum with effective temperatures 3000 K and 20\,000K 
respectively.
\label{broad}}
\end{figure}

\section*{Discussion}

The results of observations of IGR J16318-4848 in the 2--10 keV band
have shown that the spectrum of the source is unique.

\begin{itemize}
\item 
The observed 2--10 keV spectrum consists almost entirely of emission
line photons. The position of the lines indicates that we are observing
fluorescence of neutral or weakly ionized material. 

\item
The observed photoabsorption is very high -- the line-of-sight column
density $n_H L>4\cdot 10^{23}$ см$^{-2}$ (the exact value, obtainable
from the observations, depends strongly on the assumed spectral shape
of the unabsorbed source), and thus considerably exceeds that of the
interstellar gas in the Galaxy toward the source -- $\sim 2\cdot
10^{22}$ cm$^{-2}$ (Dickey \& Lockman 1990, Dame et
al. 2001). We can therefore conclude that most of the
absorption observed in the X-ray spectrum of IGR J16318-4848 takes place in
the close vicinity of the source. 
\end{itemize}

\subsection*{Active galactic nucleus?}

It is interesting to note that similar X-ray spectra with strong
absorption and a strong fluorescence line of neutral iron are observed
from some Seyfert 2 galaxies (e.g. Moran et al. 2001). This raises a
natural question: could it be that IGR J16318-4848 is a
heavily absorbed active galactic nucleus (AGN)? In this scenario, it
would be practically unfeasible to observe the galaxy itself at the
coordinates $l=335.6$, $b=-0.45$ neither in the near infrared -- due to
the huge interstellar extinction along the Galactic plane in the
direction of the source ($A_V\sim 15$-30, Schlegel et al. 1998, Dickey
\& Lockman 1990, Dame et al. 2001), nor in the far infrared because of the
strong emission of the Galaxy (see Kraan-Korteweg \& Lahav 2000).

In the extragalactic scenario, the IGR J16318-4848 distance could not
be larger than $\sim 5$~Mpc, as follows from the absence of any
significant redshift of the iron X-ray line (the present paper, de Plaa et
al. 2003). Therefore, the luminosity of the AGN corrected for
the intrinsic absorption was less than $\sim 6\cdot 10^{41}$ erg
s$^{-1}$ at 2--10 keV during the ASCA and XMM observations, and was
likely an order of magnitude higher during the outburst detected by
INTEGRAL. It is interesting that there are already 3 known heavily absorbed
AGNs ($n_H L$ ranging from $10^{23}$ to $5\cdot 10^{24}$ cm$^{-2}$),
with similar X-ray luminosities, located within 5~Mpc from us:
Centaurus A, Circinus Galaxy and NGC 4945 (Matt et al. 2000).

However, there is a very serious argument against the AGN hypothesis:
the infrared spectrum of IGR J16318-4848 (dereddened 
by $A_V\sim 20$, see Fig. \ref{broad}) is drastically
different from observed spectra of the nuclei of heavily absorbed
Seyferts (e.g. Marconi et al. 2000). We note that the contribution
from the stellar population of the host galaxy to the infrared
spectrum shown in Fig. \ref{broad} cannot be siginificant, because the
observed source is compact (the size is less than 100 pc for a
distance of 5~Mpc). 

\subsection*{Source in a molecular cloud?}

As regards the more likely origin in our Galaxy, there are practically
no galactic sources showing such an X-ray spectrum constantly. In the
case of IGR J16318-4848, the ASCA and XMM observations, separated by
approximately 8.5 years, revealed very similar spectra and fluxes (the present
paper, Murakami et al. 2003, Schartel et al. 2003, de Plaa et al. 2003). 

The only known exception is the molecular cloud SGR B2,
whose emission is possibly the result of reprocessing of a bright
flare of Sgr A$^\ast$ -- a supermassive black hole in the center of
the Galaxy (Sunyaev et al. 1993, Murakami et al. 2000). 
However, a location of IGR J16318-4848 in a big molecular cloud
can be almost certainly ruled out given the maps in the molecular CO
line (Dame et al. 2001). Moreover, we can infer from the XMM data
(Schartel et al. 2003) that the angular size of the cloud, from which
scattered and fluorescence X-ray emission would be observable, cannot
exceed 5-6''. Therefore, if the source distance is $\sim 8$ kpc, then
the linear size of the cloud is less than $\sim 0.3$ kpc. In order to
provide the necessary absorption column, $>4\cdot 10^{23}$ 
см$^{-2}$, a uniform cloud of this size must have a density $\ga
10^6$ cm$^{-3}$, which is far greater than the typical densities of
molecular clouds in the Galaxy (e.g. Solomon et al. 1987). 

\subsection*{X-ray binary in eclipse?}

Similar photoabsorption is observed during eclipses/dips in
X-ray binaries (e.g. Church et al. 1998).

We cannot exclude in principle the possibility that both observations,
by ASCA and XMM, occured during eclipse of the X-ray source -- the
ASCA observation lasted only 4 ksec, while the XMM one 27 ksec. If
the orbital period of the IGR J16318-4848 binary is longer than 2--3
days, then that could be possible but unlikely. However, 
even in this case the almost complete absence of flux at
energies $<3$ keV is unusual (compare e.g. with the spectrum of Vela
X-1 or Cen X-3 in eclipse, Sako et al. 1999, Wojdowski \& Liedahl 2001). 

\subsection*{X-ray source in a compact envelope?}

If the ASCA and XMM observations of IGR J16318-4848 did not fall on
eclipses in an X-ray binary, then the presence of strong absorption
leads inevitably to the conclusion that the source must be enshrouded
by a dense envelope. This envelope could also provide the fluorescent
iron lines ($K_\alpha$ and $K_\beta$). 

Let us consider the IGR J16318-4848 infrared counterpart. The main problem
here is that the distance to the source is unknown and the
distribution of gas and dust close to the Galactic plane is known
fairly crudely. Therefore, the amplitude of the interstellar
extinction $A_V$ may be anywhere between 0 and $\sim 30$.

Fig. \ref{broad} shows two possible near-IR spectra that have been obtained by
dereddening the measured fluxes assuming $A_V=13$ and 20. The
dereddened spectrum can be fit fairly well by a Planck spectrum with a
temperature of $T_{\rm eff}\sim 3000$ K in the former case and with
$T_{\rm eff} \ga 6000$ in the latter. The corresponding source
luminosities (assuming isotropic emission from a spherical surface) are $\sim
2\cdot 10^{37}(D/5\,{\rm kpc})^2$ and $\sim 6\cdot 10^{38} (T_{\rm
eff}/10^4\,{\rm K})^3 (D/5\,{\rm kpc})^2$ erg s$^{-1}$, where $D$ --
is the source distance. Note that the shape of the spectrum allows us to
independently conclude that the extinction toward the source cannot be
significantly higher than $A_V\sim 20$, otherwise the inferred near-IR
spectrum would become steeper than the Rayleigh--Jeans law. 

The maximum possible value $A_V\sim 20$, consistent with
interstellar absorption, corresponds to a neutral column density of $n_H 
L\sim 3\cdot 10^{22}$ cm$^{-2}$, which is at least an order of
magnitude smaller than the absorption inferred from the X-ray spectrum
of IGR J16318-4848. This suggests that the dense cloud obscuring the
X-ray source is compact, likely of the size of the binary or smaller. 

If the emission that we are observing in the spectral bands K--R
comes from the surface of a stellar companion, then this star could be
a red giant at a distance of $\sim 4$~kpc, with an effective 
temperature of $\sim 3000$ K and a luminosity of $\sim 10^{37}$ erg
s$^{-1}$. Interestingly, such a star is the companion of the famous
black hole and X-ray transient GRS 1915+105 (Greiner et
al. 2001). However, the X-ray spectrum of GRS 1915+105 is completely
different from the heavily absorbed spectrum of IGR J16318-4848.

A more promising solution, also consistent with the measured fluxes, is a
massive optical companion characterized by $T_{\rm eff}\ga 10000$ K
and $L\sim 6\cdot 10^{38} (T_{\rm eff}/10^4\,{\rm K})^3 (D/5\,{\rm
kpc})^2$ erg s$^{-1}$. This in turn suggests two major possibilities
described below. 

First, IGR J16318-4848 could be similar to the system SS 433, in
which supercritical accretion occurs via a geometrically and optically
thick disk surrounded by a powerful wind from the disk (e.g. Fabrika
1997). Then the infrared emission from IGR J16318-4848 could be
associated with the outer regions of the disk and the supermassive
optical companion, while the wind could provide the needed X-ray
absorption. We note, however, that SS 433 is famous for its jets, from
which optically thin X-ray emission (with a lot of Doppler-shifted
lines) of hot (with a temperature of the order of $10^8$ K) plasma is
detected (Kotani et al. 1996). No such spectral signatures have been
observed from IGR J16318-4848 in observations with ASCA and XMM.

It can also be that IGR J16318-4848 is similar to GX 301-2, which is an X-ray
pulsar accreting via a powerful wind from the B supergiant Wray
977, which is characterized by $T_{\rm eff}\sim 2\cdot 10^4$~K and
$L_{\rm bol}\sim 6\cdot 10^{38}$ or $\sim 5\cdot 10^{39}$ erg s$^{-1}$ 
depending on whether GX 301-2 is at a distance of 1.8 kpc (Parkes et
al. 1980) or 5.3 kpc (Kaper et al. 1995). If IGR J16318-4848 has a
companion similar to Wray 977 and is at a distance
of $\sim $2--5 kpc, then the observed near IR spectrum could be accounted
for. Moreover, the X-ray spectrum of GX 301-2 shows strong
photoabsorption varying from $2\cdot 10^{23}$ to $2\cdot 
10^{24}$ cm$^{-2}$ during a $\sim 41$ day orbital cycle and also 
flourescent iron lines that are similar to those observed in the spectrum
of IGR J16318-4848 (Endo et al. 2002). We finally note that in systems with
massive shells (such as GX 301-2 or Cyg X-3), the optically thin
emission of this shell can provide a significant contribution to the
overall IR emission at wavelengths $>$5--10 $\mu$m (Davidson \&
Ostriker 1974, Ogley et al. 2001), and we observe a change of
the spectral slope just in this region (see Fig. \ref{broad}). 

The authors thank Rodion Burenin for his assistance in the analysis of
the infrared observations. This research has made use of data obtained
through the High Energy Astrophysics Science Archive Research Center Online
Service, provided by the NASA/Goddard Space Flight Center.
\section*{Publications}

\indent

Abramowicz M., Czerny B., Lasota J., Szuszkiewicz E., AStroph. J. {\bf
332}, 646 (1988) 

Brinkman A., Dam J., Mels W. et al., Non-thermal
and Very High Temperature Phenomena in X-ray
Astronomy, Ed. by G. C. Perola and M. Salvati (In-
stitute Astronomico, Rome), 263 (1985)

Churazov E., Gilfanov M., Forman W. Jones C., Astrophys. J. {\bf 471},
673 (1996) 

Church M.J., Balucinska-Church M., Dotani, T., Asai K.,
Astroph. J. {\bf 504}, 516 (1998) 

Courvoisier ., Walter R., Rodriguez J., Bouchet L., Lutovinov A., IAU
Circ. 8063 (2003) 

de Plaa J., den Hartog P., Kaastra J., in 't Zand J.,
Mendez M., Hermsen W., Astronomer's Telegram 119 (2003) 

Dickey J.M., Lockman F.J., Ann. Rev. Astron. Astrophys. {\bf 28}, 215 (1990)

Davidson A., Ostriker J., Astroph. J. {\bf 189}, 331 (1974) 

Dame T., Hartmann D., Thaddeus P., Astroph. J. {\bf 547}, 972 (2001) 

Endo T., Ishida M., Masai K. et al., Astroph. J. {\bf 574}, 897 (2002)

Fabrika S., Astroph. and Space Science {\bf 252}, 439 (1997)

Fitzpatrick E., PASP {\bf 111}, 63 (1999)

Foschini L., Rodriguez J., Walter R., IAU Circ. 8076 (2003)

Greiner J., Cuby J.G., McCaughrean M.J., Castro-Tirado
A.J., Mennickent R.E., Astron. Astrophys. {\bf 373}, L37

Kaper L., Lamers H., Ruymaekers E. et al., Astron.Astroph., {\bf 300}, 446 (1995)

Kotani T., Kawai N., Matsuoka M., Brinkmann W., PASJ {\bf 48}, 619 (1996)

Kraan-Korteweg R.C., Lahav O.., Astron. Astrophys. Rev. {\bf 10}, 211 (2000)

Marconi A., Oliva E., van der Werf P.P., Maiolino R.,
Schreier E.J., Macchetto F., Moorwood A.F.M., Astron. Astrophys. {\bf
357}, 24 (2000)

Matt G., Fabian A.C., Guainazzi M., Iwasawa K., Bassani L., Malaguti
G., MNRAS {\bf 318}, 173 (2000) 

Moran E., Kay L., Davis M. et al., Astroph. J. {\bf 556}, 75L (2001)

Murakami H. , Koyama K. , Sakano M., Tsujimoto M., Astroph.J. {\bf
534}, 283 (2000) 

Murakami H., Dotani T., Wijnands R., IAU Circ. 8070 (2003)

Ogley R., Bell Burnell S., Fender R., MNRAS {\bf 322}, 177 (2001)

Paczynsky B., Wiita P. J., Astroph. Astroph. {\bf
88}, 23 (1980) 

Parkes G., Culhane J., Mason K., Murdin P., MNRAS {\bf 191}, 547 (1980)

Paul J. et al., Advances In Space Research, {\bf 11}, 289 (1991) 

Price S., Egan M., Carey S. et al., Astron. J. {\bf 121},
2819 (2001) 

Revnivtsev M., Gilfanov M., Churazov E., Sunyaev R., Astron. \&
Astroph., {\bf 391}, 1013 (2002) 

Sako М., Liedahl D., Kahn S, Paerels F., Astroph.J. {\bf 525}, 921 (1999) 

Schartel N., Ehle M., Breitfellner M. et al., IAU Circ. 8072 (2003)

Schlegel D.J., Finkbeiner D.P., Davis M. et al., Astrophys. J. {\bf
500}, 525 (1998) 

Schmitt et al., Astron.J {\bf 114}, 592 (1997)

Solomon P., Rivolo A., Barret J., Yahil A., Astroph. J. {\bf 319}, 730 (1987) 

Sunyaev R., Markevitch M., Pavlinsky M., Astroph. J. {\bf 407}, 606 (1993) 

Tanaka Y., Inoue H., Holt S., PASJ {\bf 46}, 37 (1994)

Trudolyubov S., Gilfanov M., Churazov E. et al., Astron. Astroph. {\bf
334}, 895 (1998)

Wojdowski P., Liedahl D., Astroph.J. {\bf 547}, 973 (2001)

\end{document}